\documentclass[rnote]{aa} 
\usepackage{amsmath}
\usepackage{aas_macros}
\usepackage{subfigure}
\usepackage{graphicx}
\usepackage{txfonts}
\usepackage{subfigure}
\usepackage{natbib}
\newcommand{\source}{\mbox{1 ES 1218+30.4 }}
\begin{document}
   \title{Modelling the variability of 1ES1218+30.4}
   \titlerunning{Modelling the variability of 1ES1218+30.4}

   \author{M.~Weidinger
          \inst{1}
          \and
          F.~Spanier
          \inst{1}
          }

   \offprints{F. Spanier, \\ \email{fspanier@astro.uni-wuerzburg.de}}
\keywords{galaxies: jets - relativistic processes - radiation mechanisms: non-thermal - BL Lacertae objects: individual: \source - galaxies: active}
   \institute{Lehrstuhl f\"ur Astronomie, University of W\"urzburg,
             Am Hubland, D-97074 W\"urzburg}
   \date{Received 22 February 2010 / Accepted 6 April 2010}
  \abstract{The blazar \source has been previously detected by the VERITAS and MAGIC telescopes in the very high energies. The new detection of VERITAS from December 2008 to April 2009 proves that \source is not static, but shows short-time variability.}{We show that the time variability may be explained in the context of a self-consistent synchrotron-self Compton model, while the long time observation do not necessarily require a time-resolved treatment.}{The kinetic equations for electrons and photons in a plasma blob are solved numerically including Fermi acceleration for electrons as well as synchrotron radiation and Compton scattering.}{The light curve observed by VERITAS can be reproduced in our model by assuming a changing level of electron injection  compared to the constant state of \source. The multiwavelength behaviour during an outburst becomes comprehensible by the model.}{The long time measurements of VERITAS are still explainable via a constant emission in the SSC context, but the short outbursts each require a time-resolved treatment.}
   \maketitle

\section{Introduction}

Blazars are a special class of active galactic nuclei (AGN) exhibiting a spectral energy distribution (SED) that is strongly dominated by nonthermal emission across a wide range of wavelengths, from radio waves to gamma rays,
and rapid, large-amplitude variability. The source of this emission is presumably the relativistic jet emitted at a narrow angle to the line of sight to the observer.\\In high-peaked BL Lac objects (HBLs) the SED shows a double hump structure as the most notable feature with the first hump in the UV- to X-ray regime and the second hump in the gamma-ray regime. Indeed, a substantial fraction of the known nearby HBLs have already been discovered with Cherenkov telescopes like H.E.S.S., MAGIC or VERITAS. The origin of the first hump is mostly undisputed: nonthermal, relativistic electrons in the jet are emitting synchrotron radiation. The origin of the second hump is still controversially debated. Up to now two kinds of models are discussed: leptonic \citep[e.g.][]{maraschi92} and hadronic \citep[e.g.][]{mannheim93} ones, which are mostly applied for other subclasses of blazars.\\Another important feature of AGNs in general and HBLs in particular is their strong variability. The dynamical timescale may range from minutes to years. This requires complex models, which obviously have to include time dependence, but this gives us also the chance to understand the mechanisms that drive AGNs. We will apply a self-consistent leptonic model to new data observed for the source \source, because those are the ones favoured for HBLs.\\\\The source HBL \source has been discovered as a candidate BL Lac object on the basis of its X-ray emission and has been identified with the X-ray source \mbox{2A 1219+30.5} \citep{wilson79,ledden81}. For the first time, \source  has been observed at VHE energies using the MAGIC telescope in January 2005 \citep{magic1218} and later from VERITAS \citep{veritas1218}. Coverage of the optical/X-ray regime is provided by BeppoSAX \citep{beppo05} and SWIFT \citep{swift07}, unfortunately the data are not always simultaneous. During the observations from December 2008 to April 2009 VERITAS also observed \source showing a time-variability \citep{veritas2010}. The observations from the MAGIC telescope have previously been modelled by \citet{michl1218}. \citet{veritas2010} claim that their new observations exhibiting variability challenge the previous models. We will show that a timedependent model using a self-consistent treatment of electron acceleration is able to model the new VERITAS data.\\\\We present the kinetic equation, which we solve numerically, describing the synchrotron-self Compton emission (Sect. \ref{sec:model}). In Sect.~\ref{sec:results} we apply our code to \source, taking the VERITAS data into account and give a set of physical parameters for the most acceptable fit. Finally, we discuss our results in the light of particle acceleration theory and the multiwavelength features.

\section{Model}
\label{sec:model}
Here we will give a brief description of the model used, for a complete overview see \citep{weidinger2010a, weidinger2010b}.\\We start with the relativistic Vlasov equation  \citep[see e.g.][]{schlick02} in the one dimensional diffusion approximation \citep[e.g.][]{schlickeiser84}, here the relativistic approximation $p\approx \gamma m c$ is used. This kinetic equation will then be solved time-dependently in two spatially different zones, the smaller acceleration zone and the radiation zone, which are assumed to be spherical and homogeneous. Both contain isotropically distributed electrons and a randomly oriented magnetic field as common for these models. All calculations are made in the rest frame of the blob.\\Electrons entering the acceleration zone (radius $R_{\text{acc}}$) from the upstream of the jet are continuously accelerated through diffusive shock acceleration. This extends the model of \citet{kirk98} with a stochastic part. The energy gain due to the acceleration is balanced by radiative (synchrotron) and escape losses, the latter scaling with $t_{\text{esc}}= \eta R_{\text{acc}}/c$ with $\eta=10$ as an empirical factor reflecting the diffusive nature of particle loss. Escaping electrons completely enter the radiation zone (radius $R_{\text{rad}}$) downstream of the acceleration zone.\\Here the electrons are suffering synchrotron losses as in the acceleration zone and also inverse-Compton losses, but they do not undergo acceleration. Pair production and other contributions do not alter the SED in typical SSC conditions and are neglected \citep{boettcher02}. The SED in the observer's frame is calculated by boosting the selfconsistently calculated photons towards the observer's frame and correcting for the redshift $z$: $I_{\nu_{\text{obs}}} = \delta^3 h\nu_{\text{obs}}/(4\pi)N_{\text{ph}}$
with $\nu_{\text{obs}}=\delta/(1+z) \nu$. The acceleration zone will have no contribution to $I_{\text{obs}}$ directly, due to the $R_{\text{i}}^2$ dependence of the observed flux at a distance $r$ ($F_{\nu_{\text{obs}}}(r) = \pi I_{\nu_{\text{obs}}}R_{\text{rad}}^2r^{-2}$) and the small size of the acceleration zone. The kinetic equation in the acceleration zone is
\begin{align}
 \label{acczone}
 \frac{\partial n_e(\gamma, t)}{\partial t} = & \frac{\partial}{\partial \gamma} \left[( \beta_s \gamma^2 - t_{\text{acc}}^{-1}\gamma ) \cdot n_e(\gamma, t) \right] + \nonumber \\  &\frac{\partial}{\partial \gamma} \left[ [(a+2)t_{\text{acc}}]^{-1}\gamma^2 \frac{\partial n_e(\gamma, t)}{\partial \gamma}\right] + \nonumber \\
& + Q_0(\gamma-\gamma_0) - t_{\text{esc}}^{-1}n_e(\gamma, t)~\text{.}
\end{align}
The injected electrons at $\gamma_0$, as the blob propagates through the jet, are considered via $Q_{\text{inj}}(\gamma , t) := Q_0 \delta(\gamma - \gamma_0)$. The synchrotron losses are calculated using Eq. \eqref{synchrotronlosses}.
\begin{align}
 \label{synchrotronlosses}
 P_s(\gamma) & = \frac{1}{6 \pi} \frac{\sigma_{\text{T}}B^2}{mc}\gamma^2 = \beta_s \gamma^2
\end{align}
with the Thomson cross-section $\sigma_{\text{T}}$. The characteristic timescale for the acceleration $t_{\text{acc}} = \left(v_s^2/(4K_{||})+2v_A^2/(9K_{||}) \right)^{-1}$ of the system is found by comparing Eq. \eqref{acczone} with \citet{schlickeiser84} with the parallel spatial diffusion coefficient $K_{||}$ not depending on $\gamma$ when using the hard sphere approximation. The characteristic timescale has an additional factor ($\propto v_A^2$) arising from the Fermi-II processes compared to shock acceleration by itself. The stochastic part of the acceleration also gives rise to the second row in Eq. \eqref{acczone}, while the first row mainly depends on Fermi-I processes. This dependence of $t_{\text{acc}}$ is important for the interpretation of the resulting electron spectra, e.g. of their slopes (depending on $t_{\text{acc}}/t_{\text{esc}}$) or the maximum energies (depending on $1/(t_{\text{acc}}\beta_s)$), see \citet{weidinger2010a} for details. For modelling SEDs and lightcurves it is primary important to ensure sensible values for $t_{\text{acc}}$. Unlike in \citet{drury99}, the energy-dependence of the escape losses is also neglected because we do not expect a pileup as suggested in \citet{schlickeiser84} at typical SSC conditions. $v_s, v_A$ are the shock and Alfv\'en speed respectively. Hence $a$ in Eq. \eqref{acczone} measures the efficiency of the shock acceleration compared to stochastic processes. Setting $v_A = 0$, i.e. $a \rightarrow \infty$, will result in a shock-only model like \citet{kirk98}.\\\\This model takes account of a much more confined shock region. Fermi-I acceleration will probably not occur over the whole blob when considering a real blazar but rather at a small region in the blob's front.
Neglecting acceleration simplifies the kinetic equation in the radiation zone to
\begin{align}
 \label{radzone1}
 \frac{\partial N_e(\gamma, t)}{\partial t}  = & \frac{\partial}{\partial \gamma}\left[\left(\beta_s \gamma^2 + P_{\text{IC}}(\gamma)\right) \cdot N_e(\gamma, t) \right] \nonumber\\&- \frac{N_e(\gamma, t)}{t_{\text{rad,esc}}} + \left(\frac{R_{\text{acc}}}{R_{\text{rad}}} \right)^3\frac{n_e(\gamma, t)}{t_{\text{esc}}}~\text{.}
\end{align}
$P_{\text{IC}}$ accounts for the inverse-Compton losses of the electrons additionally occurring (beside the synchrotron losses) \citep[e.g.][]{schlick02}:
\begin{align}
 \label{iclosses}
 P_{\text{IC}}(\gamma) & = m^3c^7h \int_{0}^{\alpha_{max}}{d\alpha \alpha \int_0^{\infty}{d\alpha_1 N_{\text{ph}}(\alpha_1) \frac{dN(\gamma,\alpha_1)}{dtd\alpha}}}~\text{.}
\end{align}
The photon energies are rewritten in terms of the electron rest mass, $h \nu = \alpha m c^2$ for the scattered photons and $h \nu = \alpha_1 m c^2$ for the target photons respectively. Equation \eqref{iclosses} is solved numerically using the full Klein-Nishina cross-section for a single electron scattering off a photon field \citep[see e.g.][]{jones68}. Here $\alpha_{max}$ accounts for the kinematic restrictions on IC scattering. In analogy to the acceleration zone the catastrophic losses are considered via $t_{\text{esc,rad}} = \eta R_{\text{rad}}/c$ with $\eta = 10$.  $t_{\text{esc,rad}}$ is the responding timescale of the electron system, which is proportional to the variability timescale in the observer's frame \citep[see e.g.][]{var95}:
\begin{align}
 \label{observersframe}
 t_{\text{var}} \propto \frac{t_{\text{esc,rad}}}{\delta}~.
\end{align}
To determine the time-dependent model SED of blazars the partial differential equation for the differential photon number density has to be solved time-dependently, which can be done numerically. The PDE \eqref{radzone2} can be obtained from the radiative transfer equation making use of the isotropy of the blob
\begin{align}
 \label{radzone2}
 \frac{\partial N_{\text{ph}}(\nu, t)}{\partial t} & = R_s - c \alpha_{\nu} N_{\text{ph}}(\nu, t) + R_c - \frac{N_{\text{ph}}(\nu, t)}{t_{\text{ph,esc}}}~\text{,}
\end{align}
where $R_s$ and $R_c$ are the production rates for synchrotron photon and the inverse-Compton respectively. $R_s$ is calculated using the well known Melrose approximation and the inverse-Compton production rate $R_c$ is treated in the most exact way, i.e. using the full Klein-Nishina cross section, see \citet{weidinger2010a}. Below a critical energy the obtained spectrum is self-absorbed due to synchrotron self-absorption, which is described by $\alpha_{\nu}$ \citep{weidinger2010a, michl1218}. The photon-loss rate is set to be the light-crossing time.

\section{Results}
\label{sec:results}
Using the parameters summed up in Table \ref{tab:1218} we were able to fit the emission of \source as a steady state with our SSC model, see Fig. 1. We used all the archival data from BeppoSAX, SWIFT in the X-ray band and the MAGIC 2006, VERITAS 2009 as well as the new released VERITAS 2010 data in the VHE to model the SED of \source \citep{beppo05, swift07, magic1218, veritas1218, veritas2010}. The derived SED is absorbed in the VHE using the EBL model of \citet{primack05} for the corresponding redshift of \source.\\The parameters of our SSC model are well winthin the standard SSC parameter region with an equipartition parameter of $0.02$. Even though PIC and MHD simulations suggest a higher magnetic field compared to particle energy (in the range of 0.1), this is a common assumption in SSC models, but has to be kept in mind with regard e.g. the stability of the blob. If one wishes to enforce higher equipartition parameters one could to use the model of \citet{nlschlick}. In order to allow strong shocks to form $v_A < v_S$ must be fulfilled, which is the case for $a=10$.\\\\Due to relatively small deviation (within the error margins) between the MAGIC 2005/VERITAS 2008 and the averaged VHE data from the VERITAS 2009 campaign we find a steady state the most plausible way to model the emission, i.e. the small fluctuations (see the overall lightcurve in \citet{veritas2010}) are not contributing significantly to the averaged observed SEDs. In the Fermi LAT energy regime our model yields a photon index of $\alpha_{\text{Fer}} = -1.69$, whichagrees well with the Fermi measurement of $-1.63\pm0.12$ \citep{fermiindex}.
\begin{table}[h]
 \caption{Model parameters for the low-state SED, basis to model the outburst by varying $Q_0$.}
 \vskip4mm
 \centering
\begin{tabular*}{1\linewidth}{@{\extracolsep{\fill}}ccccccc}
\hline
\hline
$Q_0 (\text{cm}^{-3})$  & $B(\text{G})$ & $R_{\text{acc}}(\text{cm})$ & $R_{\text{rad}}(\text{cm})$ & $t_{\text{acc}}/t_{\text{esc}}$ & $a$ & $\delta$\\
\hline
\rule[-6pt]{0pt}{21pt}   $6.25 \cdot 10^{4}$ & $0.12$ & $6.0 \cdot 10^{14}$ & $3.0 \cdot 10^{15}$ & $1.11$ & $10$ & $44$\\
\hline
\hline
\end{tabular*}
\label{tab:1218}
\end{table}
\\\\The lightcurve of \citet{veritas2010} shows a relatively strong outburst at $\approx$ MJD54861. Starting with the steady state emission (solid line, Fig. 1; parameters: Table \ref{tab:1218}) we injected more electrons $Q_0$ into the emission region at low $\gamma_0 \approx 3$. As the blob evolves in time the emission in the model at higher energies rises and drops off again when the injected electrons finally relax to the initial $Q_0$. This process can be explained as density fluctuations along the jet axis and finally fits the flare.
\begin{figure}[t]
 \vspace*{2mm}
 \begin{center}
 \includegraphics[width=8.3cm]{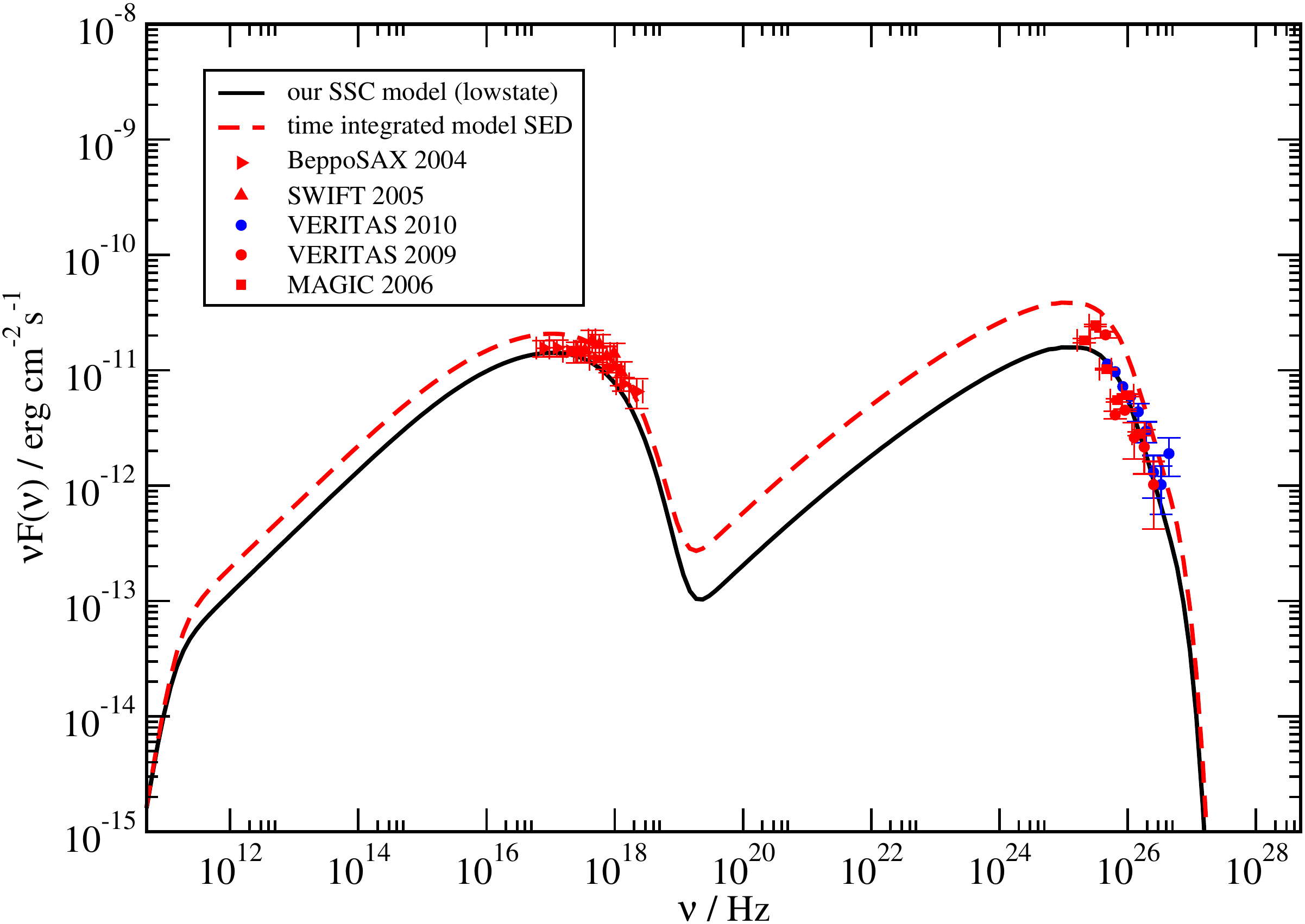}
 \end{center}
 \caption{Model SED of \source (black solid line) as derived using the described model (see Sect. \ref{sec:model}) and the parameters shown in Table \ref{tab:1218}. The VHE parts of the model SEDs have been absorbed using the EBL model of \citet{primack05}. The BeppoSAX data are from \citet{beppo05}, SWIFT from \citet{swift07}, MAGIC from \citet{magic1218}, VERITAS 2009 from \citet{veritas1218} and the blue dots are the new VERITAS 2010 data from \citet{veritas2010}. The dashed red curve shows the time integrated SED over the strong outburst shown in Fig. 2, as measured by VERITAS in 2009.}
\end{figure}
\\We found that nearly doubling the injected electron number density in a $Q_0(t)=1+b(t/t_{\text{e,var}})^3$ way with a timescale $t_{\text{e,var}} \approx 1.5$ days (as measured in the observer's frame) and then decreasing them to the initial $Q_0$ in an almost linear way on the same timescale fits the strong outburst of \source. The corresponding lightcurve of the model as well as the observed one are summarized in Fig. 2 \and 3.
\begin{figure}[t]
 \vspace*{2mm}
 \begin{center}
 \includegraphics[width=8.3cm]{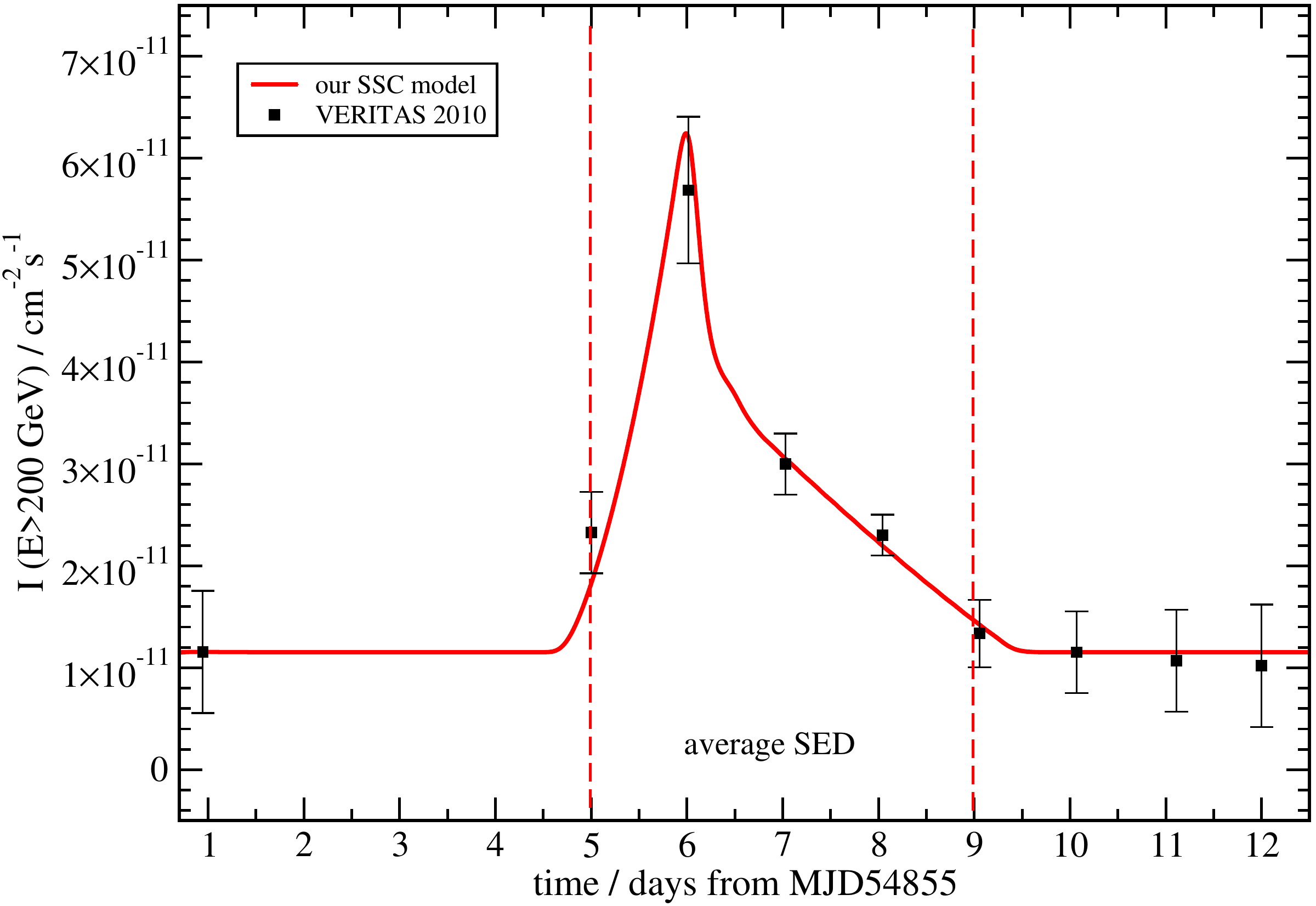}
 \end{center}
 \caption{Lightcurve of the photon flux above $200$ GeV as measured by \citet{veritas2010} (inset of their figure) in January 2009 to February 2009 and our model (red solid line). The outburst was modelled by injecting more electrons into the blob by varying $Q_0$ at a timescale of $\approx 1.5$ days (see text for details).}
\end{figure}
Figure 3 shows a more detailed view of the lightcurve in the VHE (above $200$ GeV) as well as the corresponding lightcurves in the X-Ray (between $1.2$ keV and $11$ keV) regime of BeppoSAX/SWIFT and the lower tail of the Fermi LAT energy range (between $0.2$ GeV and $22$ GeV) as predicted by our model. The latter two have been scaled down to the flux level of the VERITAS measurement, see Fig. 3, because the real fluxes are higher than the VHE flux. The model predicts the peak of the lightcurve in the Fermi regime to be $1.26$ hours ahead of the VHE one, where the X-ray regime is delayed by $0.97$ hours for \source. The delay of the X-ray band can be used to verify the model when multiwavelength data of the flaring behaviour of \source is available, while the derivation of the $0.2$ GeV to $22$ GeV lightcurve is beyond the resolution of Fermi for this source.
\begin{figure}[t]
 \vspace*{2mm}
 \begin{center}
 \includegraphics[width=8.3cm]{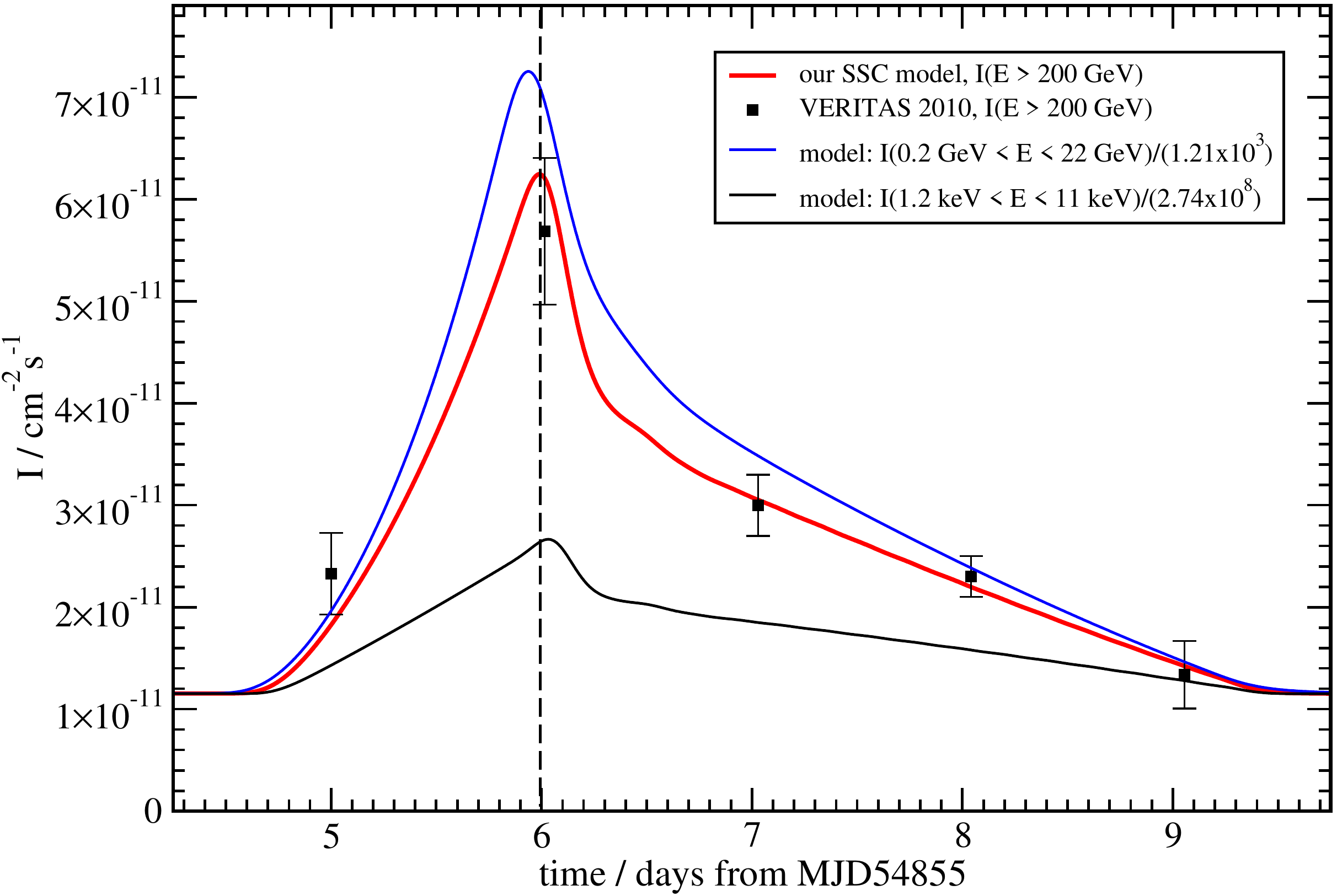}
 \end{center}
 \caption{Detailed view of the high outburst shown already in Fig. 2 as well as the behaviour of \source in the lower Fermi LAT energy band and the synchrotron regime, measurable by BeppoSAX/SWIFT during such a flare.}
\end{figure}
\\\\Additionally we plotted the time averaged SED (over the outburst from MJD54860 until MJD54864) into the SED of \source, Fig. 1 (dashed red line). As one can see only when separately considering the strongest outburst of \source within the VERITAS campaign in 2009 the aberration from a presumed steady state is significant. In contrast an average over the whole observation of \citet{veritas2010}, which is low-state most of the time, will result in a steady state as shown here or in \citet{michl1218}. For the IC photon index $\alpha$ ($\nu F_{\nu} \propto \nu^{\alpha+2}$) above $200$ GeV we get $\alpha_{\text{VHE}} = -3.53$ for the low-state (solid curve in Fig. 1), which slightly softens to $\alpha_{\text{VHE}} = -3.56$ when considering the high-state as the time-average over the single outburst shown in the lightcurve, Fig. 2 in the VERITAS range. Note that the photon index and its behaviour during an outburst in this energy range is very sensitive to the EBL absorption and thus to the EBL model used and shows a strong dependency on the considered energy range. With our model we are able to compute the spectral behaviour in the X-Ray energy range of the BeppoSAX/SWIFT satellites (i.e. $1.2$ keV $<$ E $<$ $11$ keV). The model predicts the source to be spectrally steady in this regime with a photon index $\alpha_{\text{xray}} = -2.68$ for outbursts on timescales of days. Considering shorter averaging timescales of the outburst of \source, e.g. the first or last two hours, two hours around the peak in the lightcurve, the maximum derivation from $\alpha_{\text{xray}}$ $(\alpha_{\text{VHE}})$ predicted by the model is $\pm 0.05$ $(-0.07)$, which could not be measured with current experiments and thus is considered as spectrally steady in this case.

\section{Discussion}
\label{sec:discussion}
Our results clearly show that the latest observations from the VERITAS telescope for \source still agree with a constant (steady state) emission from a SSC model when averaged over a long observation period. This is due to the relatively moderate variability of \source compared to the observation time.\\The variability may be well explained in the context of the self-consistent treatment of acceleration of electrons in the jet. We are aware that an outburst of the timescale of roughly five days as measured from \source does not necessarily require a shock in jet model, which scales down to a few minutes depending on the SSC parameters \citep{weidinger2010b}, but may also be explained as e.g. different accretion states. Nevertheless the fundamental statement remains the same: long time observation of slightly variable blazars will result in a steady state emission, while an average over a single outburst will, of course, result in a significantly different SED for the source. We are not yet able to rule out different emission models or even complex geometries of the emitting region. But we are able to model the influence of short outbursts of a source on the SED and the lightcurves in the different energy bands selfconsistently.\\
The VERITAS collaboration only shows an integrated spectrum for \source, which is due to the low flux of the source and the photon index behaviour of the combined high-states. This integrated spectrum does not show strong variations with regard to the known low-state observed by MAGIC. Our model now predicts a clear change in the spectrum, which is indicated by the dashed line in Fig. 1, which shows the average over one outburst with a slight, currently not detectable spectral softening in the VHE range, while the synchrotron peak in the BeppoSAX/SWIFT regime remains spectrally constant. This situation changes for shorter and/or stronger outbursts of an overall timescale of hours, which will result in measurable spectral evolutions in all energy regimes when considered with the presented model. Furthermore the time-resolved SEDs during a flare are comprehensible with our model. Hence with better time-resolved spectra or/and better multiwavelength coverage it should be possible to prove this model, and if the model is indeed applicable it will be a good tool to investigate the whole SED during an outburst without having all energy regimes observationally covered.\\\textit{Acknowledgments} MW wants to thank the Elitenetzwerk Bayern and GK1147 for their support. FS acknowledges support from the DFG through grant SP 1124/1.

\bibliographystyle{aa}
\bibliography{14299}

\end{document}